\documentclass[conference]{IEEEtran}
\IEEEoverridecommandlockouts

\usepackage{cite}
\usepackage{amsmath,amssymb,amsfonts}
\usepackage{algorithmic}
\usepackage{graphicx}
\usepackage{textcomp}
\usepackage{xcolor}
\usepackage{url}
\def\BibTeX{{\rm B\kern-.05em{\sc i\kern-.025em b}\kern-.08em
    T\kern-.1667em\lower.7ex\hbox{E}\kern-.125emX}}
\begin{document}

\title{Security and Privacy in the Musical Metaverse: Threat Analysis and Design Implications\\
\thanks{This work was supported by the MUSMET project funded by the EIC Pathfinder Open scheme of the European Union (grant agreement n. 101184379). Views and opinions expressed are however those of the authors only and do not necessarily reflect those of the European Union or the European Innovation Council. Neither the European Union nor the European Innovation Council can be held responsible for them.}
}

\author{\IEEEauthorblockN{Luca Turchet}
\IEEEauthorblockA{\textit{Department of Information Engineering} \\
\textit{and Computer Science} \\
\textit{University of Trento}\\
Trento, Italy \\
luca.turchet@unitn.it}
\and
\IEEEauthorblockN{Michał Kłosiński}
\IEEEauthorblockA{\textit{R\&D department} \\
\textit{7bulls.com}\\
Warszawa, Poland \\
mklosinski@7bulls.com}
}

\maketitle

\begin{abstract}
The Musical Metaverse (MM) introduces immersive, real-time environments for collaborative musical interaction, characterized by ultra-low-latency constraints, continuous multimodal data streams, and heterogeneous devices. These properties create a distinctive security and privacy landscape that differs significantly from conventional XR or multimedia systems. This paper presents a multi-layer threat analysis of MM ecosystems, identifying key assets including live musical content, expressive interaction data, identity and session metadata, and intellectual property. Threats are analyzed across network, application, data/AI, device, intellectual property rights, and social layers, with particular attention to risks arising from expressive and neurophysiological data, which enable inference, re-identification, and potential privacy violations. We describe a stakeholder-driven survey involving 14 participants from 13 organizations, revealing that neurophysiological data leakage and real-time stream disruption are perceived as the most critical risks, followed by intellectual property infringement and avatar impersonation. We further evaluate the suitability of existing security protocols under strict latency constraints, showing that conventional approaches such as TLS over TCP are often incompatible with real-time musical interaction, while lightweight, stream-oriented mechanisms (e.g., SRTP, DTLS) provide a more suitable balance between security and performance. Based on these findings, we derive a set of design guidelines for MM systems, emphasizing latency-aware security, differentiation of interaction paths, data minimization, and edge-centric processing. The results support a security-by-design approach that enables trust and compliance without compromising real-time performance.
\end{abstract}


\section{Introduction} 

The emergence of the Metaverse marks a significant shift in the way individuals interact, collaborate, and create within shared virtual environments. Within this broader landscape, the Musical Metaverse (MM) represents a domain with particularly stringent requirements for real-time interaction, expressive communication, and secure exchange of multimodal information \cite{turchet2023musical}. Musical activities, especially collaborative performance \cite{park2024research}, are characterized by extreme sensitivity to latency, jitter, synchronization accuracy, and data integrity, far exceeding those of typical social XR applications. In such settings, delays above 30 ms can substantially impair the ability of geographically distributed musicians to perform together, while inconsistent audiovisual or haptic cues can disrupt audience immersion and compromise artistic expression \cite{rottondi2016overview,boem2025issues}. 

Current research in this domain is seeking to advance the state of the art by developing an integrated ecosystem of devices, networking techniques, and software services enabling real-time, high-fidelity, multisensory musical interactions in XR environments \cite{rinaldi2024musical,buffa2024using,turchet2025towards}. Achieving this vision requires not only technical innovation, but also a robust and comprehensive approach to security and privacy, given the unprecedented scale and sensitivity of data involved. However, while different studies have identified security and privacy issues for the general field of the Metaverse \cite{di2021metaverse,chen2022metaverse,wang2023survey,kang2023security,huang2023security}, to the best of the authors' knowledge no comprehensive investigation has been conducted on such topics for the specific domain of musical activities.

As the MM is a subfield of the Metaverse, it inherits all the issues related to privacy and security of such a general field. For instance, these include pervasive user profiling enabled by massive multimodal data collection, identity theft and impersonation through avatar misuse, data leakage and unauthorized access, network-level attacks such as eavesdropping and man-in-the-middle, as well as AI-driven threats including deepfakes and adversarial manipulation \cite{wang2023survey,kang2023security,huang2023security}. Notably, security and privacy in the Metaverse may be also related to other ethical issues, such as super-realism \cite{slater2020ethics}. Nevertheless, such issues, along with the related technical solutions, may be insufficient to inform the design of secure, privacy-preserving virtual spaces where users musically interact in a trustworthy manner.

In the MM vision described in \cite{turchet2023musical}, user interactions rely on continuous streams of audio, video, motion-tracking signals, haptic stimulation, environment sensing, and, in some scenarios, biometric or neurophysiological signals, such as EEG or affective state indicators. Such data can reveal deeply personal information, including identity traits, behavioural patterns, emotional states, or health characteristics. At the same time, MM environments enable creation, manipulation, and distribution of original artistic content, raising important concerns about intellectual property rights (IPR), integrity of creative works, and protection against unauthorized recording or reproduction. 

Furthermore, the distributed, multi-vendor, and highly heterogeneous nature of emerging Metaverse infrastructures introduces new vulnerability surfaces: insecure XR devices, compromised IoT sensor nodes, untrusted network paths, malicious avatars, and misuse of real-time data channels \cite{ali2024metaverse,wang2023survey}. Traditional security protocols, often optimized for throughput, reliability, or enterprise scenarios, are not automatically suitable for ultra-low-latency audio-first environments. As a result, the MM domain demands a deep evaluation of existing and emerging security and privacy technologies, including transport-layer encryption, authentication methods, data minimization techniques, and privacy-preserving machine learning, to determine their feasibility under the tight real-time constraints.

To bridge the knowledge gaps above, this paper pursues a twofold objective: i) to identify and analyze security and privacy threats in the Musical Metaverse, including both those inherited from generic metaverse environments and those arising from its real-time, audio-centric nature; and (ii) to define feasible measures for ensuring secure and trustworthy system operation. To this end, we first characterize the security- and privacy-relevant context of MM systems, including latency constraints, multimodal data, user interactions, and intellectual property (IP) considerations. We then synthesize domain-specific threat categories spanning network, device, identity, AI/ML, biometric, and IPR-related risks. Building on a survey conducted with 14 participants from 13 academic and industrial organizations, we further ground this analysis in stakeholder perspectives. Finally, we evaluate the suitability of existing security protocols under strict latency constraints and derive design implications for future MM ecosystems based on latency-aware, context-dependent security and privacy strategies.

\section{Related Work} 

Security and privacy in the Metaverse have been extensively studied across multiple domains, including extended reality (XR), distributed systems, multimedia networking, artificial intelligence, and blockchain-based ecosystems. Several surveys provide comprehensive overviews of the technological foundations and associated risks, highlighting how the convergence of immersive interaction, large-scale data collection, and heterogeneous infrastructures creates a complex threat landscape \cite{wang2023survey,huang2023security,ali2024metaverse,chen2022metaverse,kang2023security,di2021metaverse}.

A broad consensus emerges in the literature regarding the main categories of security and privacy threats. These include pervasive user profiling enabled by continuous multimodal sensing (e.g., motion, biometrics, and behavioral data), identity theft and avatar impersonation, unauthorized data access and leakage, and network-level attacks such as eavesdropping, man-in-the-middle, and denial-of-service \cite{wang2023survey,huang2023security,kang2023security}. Recent work has further highlighted the central role of avatars as a primary source of privacy risk, showing that avatar-based interactions enable the inference of sensitive personal attributes (e.g., identity, behavior, and emotions), as well as new forms of social threats such as harassment and impersonation, thereby exposing users to both technical and socio-behavioral privacy violations \cite{eltanbouly2025avatar}. Additional concerns arise from AI-driven threats, including deepfake generation, adversarial manipulation, and biased decision-making, as well as from blockchain-related vulnerabilities affecting digital assets and virtual economies \cite{chen2022metaverse,ali2024metaverse}. These risks are further amplified by the scale, interoperability, and decentralization of metaverse systems, which complicate trust management, governance, and accountability \cite{wang2023survey,ali2024metaverse}.

A key insight across these works is that the Metaverse inherits vulnerabilities from its underlying technologies while simultaneously introducing new ones due to its intrinsic characteristics, namely immersiveness, real-time interaction, and the tight coupling between physical, digital, and social spaces \cite{wang2023survey,di2021metaverse}. For instance, the large-scale collection of fine-grained sensory and behavioral data significantly increases the risk of privacy invasion and inference attacks, while the blending of virtual and physical environments enables new forms of cyber-physical threats affecting both digital assets and real-world safety \cite{huang2023security,ali2024metaverse}.

Within this broader context, emerging research highlights limitations of existing approaches when applied to domains characterized by strict real-time constraints and highly expressive interaction. In such settings, threats related to timing perturbations, subtle data manipulation, and continuous data streams become more critical than in conventional XR or social metaverse applications. Minor alterations in data or communication timing (often negligible in traditional systems) may significantly affect synchronization, perception, and interaction quality.

Another important gap concerns IP and content integrity \cite{gupta2024evolution}. While prior work addresses ownership and digital asset protection, it often assumes static or post-hoc content management. In contrast, interactive environments involving real-time content creation and exchange raise additional challenges, such as protecting live streams, preserving authorship, and preventing unauthorized capture or redistribution during ongoing interactions \cite{kang2023security,ali2024metaverse}.

Finally, recent studies emphasize the importance of system-level, privacy-preserving architectures beyond protocol-level security. Approaches such as data minimization, edge and local processing, federated learning, and separation between interaction and analytics pipelines are increasingly proposed to mitigate risks associated with large-scale data aggregation and long-term profiling \cite{huang2023security,ali2024metaverse,kang2023security}. This perspective reflects a shift from isolated security mechanisms toward holistic, context-aware design strategies.

In general, the analysis of the literature surveyed above suggests that while existing metaverse security frameworks provide a solid foundation, they are not fully sufficient for domains requiring ultra-low latency, continuous multimodal interaction, and high semantic sensitivity of data, such as the MM. These conditions motivate the need for more context-specific security and privacy models, particularly in emerging application areas such as real-time collaborative and creative environments.

\section{Methodology} 

Our security and privacy analysis of the MM adopted a multi-source qualitative methodology combining state-of-the-art review, contextual system analysis, stakeholder input, and the derivation of feasible mitigation measures. This approach enables a balanced assessment grounded in both established research and the practical constraints of real-time, immersive musical interaction environments.

First, a structured review of the scientific literature, technical reports, and relevant standards was conducted to identify security and privacy threats associated with the Metaverse, XR systems, IoT, and distributed multimedia infrastructures. Particular attention was given to real-time communication, low-latency networking, biometric data protection, user safety, and privacy-preserving data processing, establishing a baseline threat landscape and known countermeasures.

Second, this baseline was refined through a contextual analysis of MM characteristics, including ultra-low-latency interaction, continuous multimodal data streams, expressive performance data, XR-based collaboration, and heterogeneous devices and networks. This step highlights domain-specific risks emerging from tight timing constraints, audio-centric interaction, and the handling of sensitive behavioral or physiological data.

Third, stakeholder-driven input was incorporated through a structured survey of domain experts to capture perceived risks, validate assumptions from the literature, and prioritize threats based on practical relevance. Such a survey involved only open-ended questions.

Finally, a set of mitigation measures was derived, focusing on solutions that are compatible with performance constraints, architectural requirements, and regulatory considerations. This aimed at ensuring that the proposed approaches are not only theoretically sound but also realistically applicable in MM systems.

\subsection{Survey participants and procedure} 

The survey was administered to 14 selected experts (11 males, 3 females) in various technological areas related to the MM. They belonged to 13 organizations, namely 8 industries (Somnium Space, PatchXR, VRoom Studio, Lynx-R, Atmoky, HiFi Berry, 7bulls, GTEC) and 5 academic institutions (University of Trento, Polytechnic University of Turin, University of Applied Sciences and Arts of Southern Switzerland, National Research Council of Italy, KTH Royal Institute of Technology). The respondents included researchers, engineers, software developers, networking specialists, XR experts, audio technology researchers, and system architects, each contributing expertise aligned with the MM components developed within their respective organizations. This composition aimed at grounding the findings in both implementation and research perspectives. Specifically, all the involved participants belonged to the consortium of the MUSMET\footnote{\url{https://musmet.eu}} project \cite{turchet2025towards}, which aims at advancing the state-of-the-art of MM technologies and interaction paradigms. Each partner of the consortium is responsible to advance specific hardware and/or software components. These include 5G/edge/cloud infrastructures, XR headsets, embedded audio devices, brain-computer interfaces (BCIs), 3D audio, networked music performance systems, and MM concert platforms. 

Respondents were not asked to document already deployed controls; rather, they were instructed to anticipate likely security, privacy, and IPR challenges associated with MM components and use cases. Therefore, the survey captured perceived future risk rather than retrospective incident data. Specifically, the survey was structured as follows.
Participants first identified the types of data their components would collect, process, or transmit. These included real-time audio and audiovisual streams, 3D spatial and interaction data, identifiable user and avatar information, network and session metadata, and biometric or neurophysiological signals such as EEG and affective indicators. They then reflected on how the hardware and software components they were developing might be exposed to the categories of two widely recognized frameworks:

\begin{itemize}
\item STRIDE (Spoofing, Tampering, Repudiation, Information Disclosure, Denial of Service, Elevation of Privilege) \cite{howard2006security}, to systematically identify anticipated security threats;
\item LINDDUN (Linkability, Identifiability, Non-repudiation, Detectability, Disclosure of Information, Unawareness, Non-compliance) \cite{deng2011privacy} to explore privacy risks arising from immersive, sensor-rich environments.
\end{itemize}

This dual-framework approach ensured balanced coverage of technical, organizational, and user-centric risks, while remaining compatible with the MM distributed and heterogeneous architecture. In addition, participants were asked to answer a set of open-ended questions focusing on planned technical and organizational security measures, anticipated risks and potential misuse scenarios, the impact of security and privacy on system design choices (e.g., performance-security trade-offs), and strategies to enhance trustworthiness, user confidence, and early-stage risk mitigation in MM ecosystems.

Responses were analyzed using an inductive thematic analysis. The two authors independently reviewed the responses, identified recurring security and privacy themes, and grouped them into common categories through discussion until consensus was reached. The prioritization reported in Table II reflects the frequency with which specific risks were identified across participants together with the perceived severity expressed in their responses. Risks consistently highlighted by a large majority of participants and considered capable of significantly affecting MM operation were classified as High, whereas risks mentioned less frequently or with lower expected impact were classified as Medium.

%
%
%
%

\section{Results} 

The results can be organized into four layers: ecosystem characterization, stakeholder risk prioritization, layered threat analysis, and protocol and architecture assessment.

\subsection{Security- and privacy-relevant characteristics of the ecosystem}

The analysis begun by defining a threat model tailored to the MM. Key stakeholders include performers, composers, audiences, music teachers/students, and platform operators, each generating and consuming distinct categories of data within dynamic, multi-user environments. These interactions span both real-time performance contexts and asynchronous activities such as rehearsal, teaching, and content distribution. The primary assets to be protected are: (i) live musical content, (ii) expressive interaction data, (iii) identity and session metadata, (iv) IP, and (v) the availability and continuity of low-latency services.

Security objectives extend beyond classical confidentiality, integrity, and availability to include real-time consistency, synchronization integrity, and protection against inference from high-resolution behavioral data. In MM contexts, timing precision, spatial coherence, and interaction fidelity are themselves critical assets, as even minor disruptions may affect both user experience and artistic outcomes. Consequently, ensuring temporal integrity and consistency across distributed participants becomes a core security requirement.

Threat actors include external adversaries attempting interception or disruption, malicious or misbehaving participants engaging in impersonation or unauthorized recording, compromised or semi-trusted services within the infrastructure, and passive observers capable of inferring sensitive information from accessible data streams. Notably, privacy risks frequently arise from correlation, aggregation, and inference processes rather than from explicit data breaches, particularly when multiple data modalities (e.g., audio, motion, and physiological signals) are combined. 

A central result concerns data sensitivity stratification, summarized in Table \ref{tab:data}. Different data categories exhibit varying levels of sensitivity. Real-time audio and audiovisual streams are not only privacy-relevant but also critical for preserving artistic integrity, as their manipulation can directly impact performance quality. Expressive interaction data (such as gesture, posture, and timing) may encode behavioral signatures that enable user profiling or re-identification. Biometric and neurophysiological data (e.g., EEG, heart rate, affective signals) represent the highest level of sensitivity, as they may reveal cognitive or emotional states and require strict safeguards, purpose limitation, and controlled access. Even seemingly less sensitive metadata, such as session logs and device identifiers, can contribute to long-term tracking and linkage when aggregated.

System constraints strongly shape security design. MM systems operate under ultra-low latency (often $<$ 30 ms), continuous high-frequency data exchange, heterogeneous devices and networks, distributed processing pipelines, and dynamically evolving session compositions. These characteristics introduce unavoidable trade-offs between performance and protection. In particular, they limit the applicability of conventional, computation-heavy security mechanisms in latency-critical interaction paths and necessitate performance-aware, context-dependent protection strategies that can preserve both security and interaction quality.

\begin{table*}[htbp]
\caption{Data Sensitivity in MM Environments}
\label{tab:data}
\centering
\begin{tabular}{|l|l|c|l|}
\hline
\textbf{Data Type} & \textbf{Examples} & \textbf{Risk Level} & \textbf{Key Concerns} \\
\hline
Musical content & Audio streams, performances & High & IPR leakage, manipulation \\
\hline
Expressive data & Motion, timing, gestures & Very High & Re-identification, profiling \\
\hline
Biometric data & EEG, heart rate, affect & Critical & Irreversible privacy loss \\
\hline
Metadata & IDs, logs, sessions & Medium--High & Tracking, linkage \\
\hline
Derived data & AI outputs, profiles & High & Inference, misuse \\
\hline
\end{tabular}
\end{table*}

\subsection{Stakeholder risk prioritization}

Stakeholder input provided a complementary perspective to the technical analysis, revealing both convergence in perceived threats and a clear prioritization driven by the specific characteristics of MM environments. Overall, respondents consistently emphasized that risks affecting real-time interaction and sensitive data exposure are of greater concern than those typically prioritized in conventional distributed or XR systems.

From a security standpoint, three major threat categories emerge as particularly critical: identity spoofing, data tampering, and denial-of-service attacks. Identity spoofing is considered especially harmful because it can enable unauthorized access, undermine trust between participants, and lead to fraudulent attribution of creative works. Data tampering is also perceived as a severe risk, particularly in the MM context where even minimal alterations (such as micro-timing shifts or subtle distortions) can significantly degrade performance quality or bias adaptive systems. Denial-of-service attacks are similarly critical, as MM interactions cannot tolerate disruptions in the same way as delay-tolerant applications; even short interruptions may irreversibly affect rehearsals, performances, or collaborative sessions.

From a privacy perspective, stakeholders highlighted concerns related to linkability, identifiability, and lack of user awareness. Expressive interaction data and neurophysiological signals were frequently regarded as capable of enabling re-identification across sessions or platforms when combined with other data sources. In addition, respondents emphasized that users may not fully understand the extent of data collection and processing in immersive environments, particularly when advanced sensing technologies such as BCIs or affective computing are involved. Regulatory aspects further reinforced these concerns, as emerging frameworks (e.g., GDPR and the EU AI Act) impose strict requirements on the handling of sensitive data, especially those related to physiological or behavioral signals.

Importantly, stakeholder input allowed the derivation of a relative prioritization of risks, summarized in Table \ref{tab:risks}. The two highest-priority concerns are neurophysiological data leakage and real-time stream disruption. The former is associated with potentially irreversible privacy harm due to the exposure of highly sensitive biometric data, while the latter directly threatens the viability of MM interaction by violating strict latency and synchronization constraints. Medium-priority risks include IP infringement (through unauthorized recording or redistribution of collaborative performances) and avatar impersonation, which may lead to social engineering and reputational damage. Notably, the prominence of privacy leakage and availability-related risks over more traditional concerns (e.g., repudiation) reflects the distinctive priorities of MM systems, where interaction continuity and data sensitivity are central.

Beyond risk identification, stakeholders also highlighted several gaps and unmet needs. In particular, there was a strong demand for performance-aware security design that explicitly accounts for latency constraints, as well as for integrated identity management solutions aligned with zero-trust principles. Respondents further emphasized the need for coordinated governance across system components and clearer transparency mechanisms to inform users about active sensing and data processing. These observations reinforce the importance of adopting a holistic and context-aware approach to security and privacy in MM environments, bridging technical, organizational, and user-centric considerations.

\begin{table*}[htbp]
\caption{Stakeholder Risk Prioritization}
\label{tab:risks}
\centering
\begin{tabular}{|l|c|l|}
\hline
\textbf{Risk} & \textbf{Priority} & \textbf{Rationale} \\
\hline
Neurophysiological data leakage & High & Irreversible privacy exposure \\
\hline
Real-time stream disruption & High & Breaks interaction and synchronization \\
\hline
IPR infringement & Medium & Economic and artistic impact \\
\hline
Avatar impersonation & Medium & Trust and identity risks \\
\hline
Metadata profiling & Medium & Long-term tracking \\
\hline
\end{tabular}
\end{table*}

\subsection{Layered threat analysis}

\begin{figure}[ht!]
\centerline{\includegraphics[width=\columnwidth]{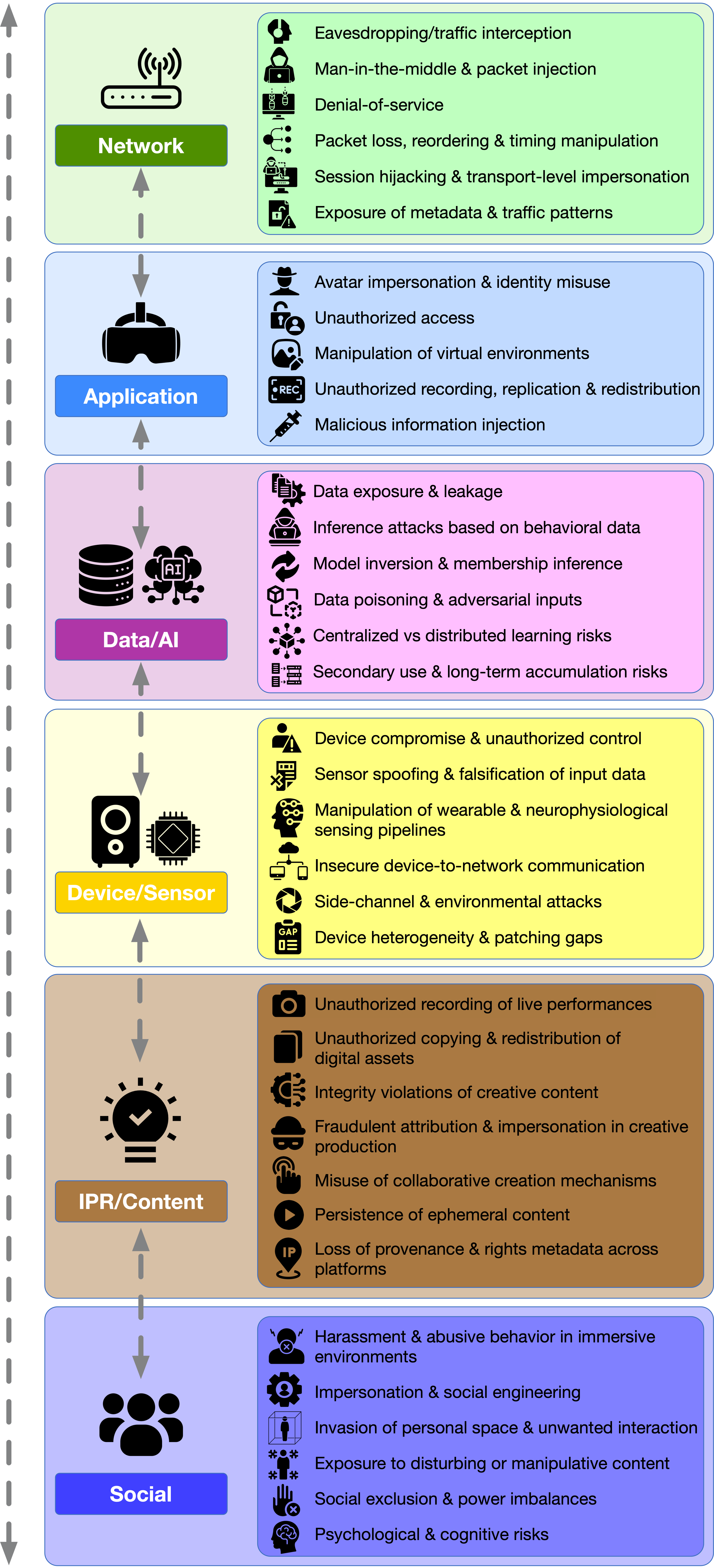}}
\caption{Layered threat model in the Musical Metaverse, illustrating how security and privacy risks span network, application, data/AI, device, intellectual property, and social layers. Arrows indicate interdependencies and cross-layer propagation of threats.}
\label{fig:layered_threat_model}
\end{figure}

The analysis of MM systems reveals a multi-layered threat landscape, where vulnerabilities arise across interconnected components spanning networking, application logic, data processing, devices, IP, and social interaction (see Fig. \ref{fig:layered_threat_model}). Rather than being isolated, these threats often propagate across layers, amplifying their impact due to the tight coupling between real-time interaction, multimodal data streams, and distributed system architectures.

\subsubsection{Network and transport layer}
At the network and transport layer, the main threats include: eavesdropping and traffic interception (where attackers may capture data packets without necessarily altering them, leading to unauthorized access to live musical performances, leakage of expressive or behavioral data, or inference of participation patterns or collaboration structures); man-in-the-middle and injection attacks (where network- and transport-layer vulnerabilities may allow attackers to intercept, modify, or inject packets into ongoing communication streams, causing degradation of audio quality or timing precision, disruption of synchronization between participants, injection of malicious or misleading control signals); packet loss, reordering, and riming manipulation (where adversaries may deliberately introduce delays or reorder packets to disrupt musical coherence without fully interrupting communication, causing the degradation of synchronization between participants, directly affecting musical coordination and perceived interaction quality); denial-of-service (where attackers overwhelm network links, transport endpoints, or session management mechanisms, causing interrupted sessions that may irreversibly disrupt live performances, with repeated disruptions potentially eroding user trust and willingness to participate); session hijacking and transport-level impersonation (allowing attackers to take control of an existing communication session or masquerade as a legitimate participant); and exposure of metadata and traffic patterns (network- and transport-layer communication may expose metadata, such as packet timing, size, frequency, and endpoint identifiers, which can reveal sensitive contextual information, leading to inference of rehearsal schedules or performance duration, identification of prominent performers or collaborators, or reconstruction of interaction intensity and group dynamics). 


\subsubsection{Application layer}
At the application layer, threats become more tightly coupled with domain semantics. These include: avatar impersonation and identity misuse (where attackers assume the identity of legitimate participants or to create misleading representations, causing unauthorized participation in musical sessions, reputational harm to performers or creators, or  fraudulent attribution of musical actions or content); unauthorized access to virtual environments such as virtual stages or rehearsal rooms (potential consequences include exposure of private rehearsals or creative processes, disruption of live performances, or leakage of sensitive interaction or performance data); manipulation of virtual environments and content (where attackers may alter parameters, inject malicious content, or subtly modify the virtual setting in ways that affect musical interaction, for instance altering spatial audio parameters or instrument settings, injecting misleading visual or auditory cues, or modifying shared scores or performance artifacts); unauthorized recording, replication, and redistribution of live performances and user-generated content (notably, the presence of built-in recording and streaming functionalities in MM platforms increases the risk of IP violations); and injection of malicious or misleading information (such as altered audio streams, misleading visual elements, or corrupted interaction data, which can mislead participants about the state of the performance, disrupt synchronization and coordination, or degrade trust in the virtual environment). Because MM interactions rely heavily on shared perception and coordinated action, application-layer attacks can be particularly difficult to detect: small alterations in spatial audio rendering, visual cues, or interaction parameters may appear as normal system variability while subtly degrading performance. 


\subsubsection{Data and AI/ML layer}
The data and AI/ML layer introduces further complexities related to both security and privacy. Threats include: data exposure and leakage across the processing pipeline (as MM data streams are continuous and high-resolution, even brief exposure events can result in significant leakage, with cumulative effects over time; leakage of expressive or performance-related data may also enable imitation, profiling, or misuse of artistic output); inference attacks based on expressive or behavioral data (whereby sensitive information is derived from musical interaction data that may appear innocuous in isolation, enabling re-identification of users across sessions or platforms, reconstruction of individual performance styles or signatures, or inference of emotional, cognitive, or physical states); model inversion (i.e., reconstructing approximate input data from trained models); membership inference (i.e., determining whether specific users or sessions contributed to training); data poisoning (i.e., malicious inputs are injected to distort model behavior, causing degraded synchronization or interaction quality, systematic bias in adaptive behavior, or subtle manipulation of collaborative dynamics); adversarial inputs (i.e., inputs crafted to trigger incorrect or unstable outputs, causing learning-based components to misinterpret interaction states or produce unintended responses);  centralized vs distributed learning risks (centralized learning approaches, where raw data are collected and stored in a central repository, create high-value targets and amplify the consequences of breaches or misuse;  distributed approaches, such as federated learning, reduce exposure of raw data, but have risks related to leakage through shared gradients or model updates, poisoning or manipulation of distributed training processes, inference attacks targeting shared model parameters); secondary use and long-term accumulation risks (MM data and models may persist beyond individual sessions, enabling long-term accumulation and secondary use; over time, aggregated datasets and evolving models may reveal patterns not apparent in short-term interaction, which creates risks related to longitudinal profiling of users or performers, unintended cross-context inference, or gradual erosion of privacy guarantees).



\subsubsection{Device and sensor layer}

At the device and sensor layer, the main threats include: device compromise and unauthorized control (where attackers gain access to XR headsets, embedded audio devices, or connected sensors, allowing them to intercept, modify, or inject data streams, leading to leakage of sensitive interaction data, disruption of user perception, or manipulation of musical outputs); sensor spoofing and falsification of input data (where adversaries inject fabricated or manipulated signals into sensing pipelines, such as motion tracking or neurophysiological sensors, causing incorrect interpretation of user actions, distortion of expressive performance data, or misalignment between physical and virtual interaction states); manipulation of wearable and neurophysiological sensing pipelines (where attackers exploit vulnerabilities in BCI or affective sensing systems to alter or infer sensitive cognitive or emotional information, leading to privacy violations, unintended behavioral influence, or misuse of highly sensitive biometric data); insecure device-to-network communication (where insufficiently protected interfaces between devices and processing nodes may expose data to interception or tampering, resulting in leakage of raw sensor data, unauthorized access to device-level information, or degradation of communication integrity); side-channel and environmental attacks (where attackers infer sensitive information from indirect signals such as power consumption, timing patterns, or physical surroundings, enabling reconstruction of user behavior or interaction patterns without direct access to primary data streams); and device heterogeneity and patching gaps (where the coexistence of diverse hardware and firmware configurations leads to inconsistent security levels, increasing the attack surface and enabling exploitation of unpatched or poorly secured components).


\subsubsection{IPR and content layer}

At the IPR and content layer, the main threats include: unauthorized recording of live performances (where adversaries capture audio, video, or multimodal interaction streams without consent, leading to uncontrolled dissemination of creative works and loss of exclusivity for performers and creators); unauthorized copying and redistribution of digital assets (where musical content, virtual instruments, or performance artifacts are duplicated and shared across platforms, causing economic loss and undermining ownership rights); integrity violations of creative content (where attackers modify musical data, performance parameters, or shared artifacts, leading to altered or corrupted artistic outputs, misrepresentation of creative intent, or degradation of performance authenticity); fraudulent attribution and impersonation in creative production (where malicious actors claim authorship or contribution to musical content, resulting in reputational damage and disputes over ownership or royalties); misuse of collaborative creation mechanisms (where shared environments and co-creation tools are exploited to extract, replicate, or manipulate contributions from other participants, enabling unauthorized reuse or appropriation of creative material); persistence of ephemeral content (where data intended to be transient, such as live performances or rehearsals, are retained through logging, caching, or recording mechanisms, leading to unintended long-term availability and increased exposure to misuse); and loss of provenance and rights metadata across platforms (where content transferred between systems or MM platforms may lose associated ownership or licensing information, enabling unauthorized reuse and weakening enforceability of IP protections).

\subsubsection{Social and user safety layer}

At the social and user safety layer, the main threats include: harassment and abusive behavior in immersive environments (where users are subjected to verbal, visual, or behavioral aggression, amplified by the sense of presence in XR, potentially causing psychological distress and discouraging participation); impersonation and social engineering (where attackers exploit identity ambiguity or avatar-based interactions to deceive users, manipulate trust relationships, or extract sensitive information); invasion of personal space and unwanted interaction (where virtual proximity, gestures, or simulated physical contact are used inappropriately, leading to discomfort or perceived violation of personal boundaries); exposure to disturbing or manipulative content (where users encounter misleading, harmful, or emotionally triggering content, which may influence behavior, perception, or decision-making); social exclusion and power imbalances (where certain users or groups are marginalized, excluded from interactions, or subjected to unequal treatment, affecting fairness and inclusivity in collaborative environments); and psychological and cognitive risks (where prolonged exposure to immersive interaction, combined with manipulation or stress-inducing stimuli, may affect users’ mental well-being, perception, or decision-making processes, particularly in highly realistic or emotionally engaging scenarios).

\medskip
The layered nature of these threats is summarized in Table \ref{tab:layers}, which illustrates how different attack vectors map onto system components and highlights their specific implications in MM environments. Overall, the results demonstrate that effective protection requires a holistic, cross-layer approach that accounts for the interplay between timing constraints, data sensitivity, and interaction semantics.

\begin{table*}[htbp]
\caption{Layered Threat Model in MM Systems}
\label{tab:layers}
\centering
\begin{tabular}{|l|l|l|}
\hline
\textbf{Layer} & \textbf{Main Threats} & \textbf{MM-Specific Impact} \\
\hline
Network & Man-in-the-middle, denial-of-service, timing attacks & Loss of synchronization \\
\hline
Application & Impersonation, manipulation & Degraded interaction semantics \\
\hline
Data / AI & Inference, poisoning & Re-identification, bias \\
\hline
Device & Spoofing, compromise & Physical--virtual inconsistency \\
\hline
IPR & Recording, copying & Loss of authorship \\
\hline
Social & Harassment, deception & Reduced trust and safety \\
\hline
\end{tabular}
\end{table*}

\subsection{Protocol and architecture assessment}

\begin{figure*}[ht!]
\centerline{\includegraphics[width=0.7\textwidth]{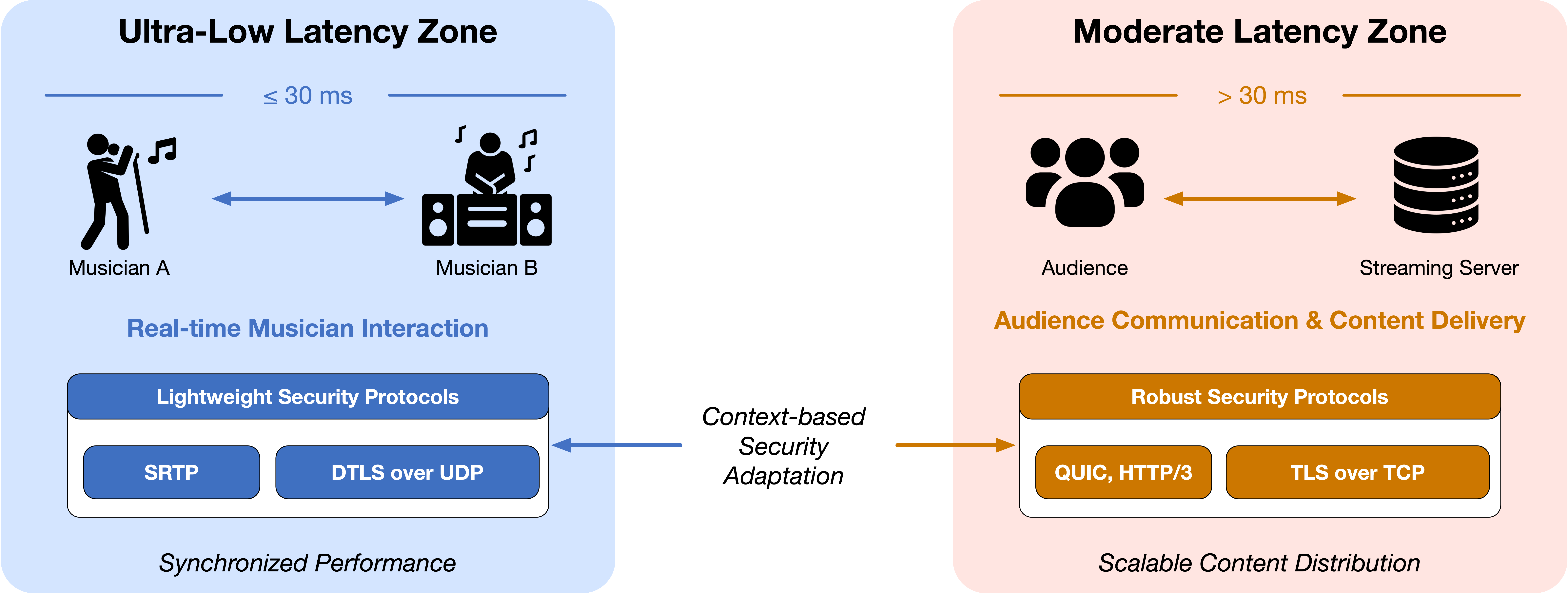}}
\caption{Latency-aware security model in the Musical Metaverse. Ultra-low-latency interaction paths (e.g., musician-to-musician) require lightweight, stream-oriented protocols such as SRTP and DTLS, while less time-critical paths (e.g., audience-facing communication) can leverage more robust protocols such as QUIC/HTTP3 and TLS.}
\label{fig:latency}
\end{figure*}

The protocol and architectural analysis highlights a key finding of this study: security mechanisms in MM systems must be adapted to latency constraints and interaction contexts. MM environments comprise at least two distinct latency zones, namely ultra-low-latency interaction paths (e.g., musician-to-musician communication) and less time-critical paths (e.g., musician-to-audience or audience-to-audience communication). This distinction directly influences protocol selection, as mechanisms suitable for one context may be incompatible with the performance requirements of another (see Fig. \ref{fig:latency}).

For latency-critical interaction paths, security solutions must preserve strict timing constraints, typically below 30 ms end-to-end delay \cite{rottondi2016overview}, while ensuring deterministic behavior and minimal jitter. These requirements limit the applicability of transport protocols relying on retransmission or buffering. In contrast, less time-sensitive communication paths can tolerate higher latency and variability, enabling the use of more computationally intensive protection mechanisms. This differentiation reinforces the need for a latency-aware security model that adapts protection strategies to each communication flow.

The evaluation of existing protocols reveals a trade-off between security strength and performance. As summarized in Table \ref{tab:protocols}, TLS over TCP, despite its robust encryption and widespread adoption, is generally unsuitable for ultra-low-latency MM interaction because of head-of-line blocking and retransmission delays. DTLS over UDP provides a more suitable alternative by avoiding TCP’s blocking behavior while maintaining comparable security guarantees, although handshake overhead and key management require careful control. SRTP is particularly well suited to MM, offering lightweight, stream-oriented protection with predictable per-packet overhead and compatibility with continuous media streams.

Protocols such as QUIC and HTTP/3 offer advantages in scalability, congestion control, and robustness, making them suitable for audience-facing communication and content distribution. However, their complexity and latency variability make them less appropriate for real-time interaction. Lightweight scrambling techniques can complement cryptographic protection in large-scale dissemination scenarios but cannot replace end-to-end security for sensitive interaction data.

A key conclusion of this analysis is that no single protocol or security mechanism is sufficient across all MM scenarios. Instead, hybrid approaches are required, combining different protocols and techniques depending on latency constraints, data sensitivity, and interaction context. For example, musician-to-musician communication may rely on SRTP combined with carefully managed DTLS, while audience-facing services may integrate QUIC-based delivery with additional access control mechanisms.

At the architectural level, the results point to design principles extending beyond protocol selection. System architectures should separate latency-critical interaction paths from control, signaling, and analytics components to enable differentiated security strategies. Sensitive data, particularly expressive and biometric streams, should be processed locally or at the edge to reduce exposure and latency, while data minimization and purpose limitation should mitigate aggregation and long-term profiling risks. Finally, distributed and federated processing can balance performance, scalability, and privacy in heterogeneous MM ecosystems.

The interplay between protocol selection and architectural design is summarized in Table \ref{tab:protocols}, which maps security mechanisms to latency zones and application contexts. Overall, the findings demonstrate that effective security in MM systems requires a co-design approach, where communication protocols, system architecture, and application requirements are jointly considered to achieve both protection and performance.

\begin{table*}[htbp]
\caption{Protocol suitability across MM latency zones}
\label{tab:protocols}
\centering
\begin{tabular}{|l|c|c|c|}
\hline
\textbf{Protocol} & \textbf{Latency Impact} & \textbf{Security Level} & \textbf{MM Suitability} \\
\hline
TLS/TCP & High & Strong & Low (non-real-time only) \\
\hline
DTLS/UDP & Low & Strong & High (real-time capable) \\
\hline
SRTP & Very Low & Strong & Very High (media streams) \\
\hline
QUIC / HTTP/3 & Medium & Strong & Medium (audience-facing) \\
\hline
Lightweight Scrambling & Very Low & Limited & Complementary only \\
\hline
IPsec / VPN & Medium--High & Strong & Low (infrastructure-level) \\
\hline
\end{tabular}
\end{table*}

\section{Discussion} 

Our analysis supports a clear claim: the MM is a setting in which security, privacy, timing, and artistic integrity are tightly coupled. Several implications follow.

First, timing sensitivity fundamentally reshapes the notion of attack severity. In conventional distributed systems, disruptions are evaluated in terms of downtime, throughput degradation, or explicit data corruption. In MM environments, however, even subtle timing perturbations (such as jitter, micro-delays, or minor distortions) can degrade synchronization and impair musical coordination without triggering conventional fault detection mechanisms. As a result, threat models must account for low-amplitude, high-impact attacks that operate below traditional detection thresholds while still compromising interaction quality.

Second, data sensitivity extends beyond traditional personal identifiers. The results show that expressive interaction data (such as timing, gesture, posture, and performance dynamics) can act as quasi-biometric signals, enabling re-identification and behavioral inference. This is further amplified by biometric and neurophysiological data, which introduce risks of irreversible privacy exposure. Importantly, these data types are not auxiliary but integral to system functionality. Consequently, MM systems challenge conventional data classification approaches, requiring context-dependent and multi-layered protection strategies rather than uniform policies.

Third, IPR protection emerges as a core technical concern rather than a purely legal one. In MM environments, creative content is generated, transformed, and exchanged in real time. Unauthorized recording, subtle manipulation, or loss of provenance directly impact both artistic integrity and economic value. This tight coupling between security and creative processes distinguishes MM systems from other metaverse applications and necessitates integrated protection mechanisms operating during interaction, not only post hoc.

Fourth, the results highlight the importance of layered and context-aware security design. Threats span multiple layers (network, application, data/AI, device, IPR, and social) and cannot be addressed effectively through isolated mechanisms. Moreover, system constraints such as ultra-low latency, continuous data streams, and heterogeneous infrastructures limit the applicability of conventional security approaches. This reinforces the need for solutions that are both technically robust and performance-compatible.

Fifth, security must be differentiated across interaction contexts. The distinction between ultra-low-latency (e.g., musician-to-musician) and less time-critical (e.g., audience-facing) communication paths implies that protection mechanisms cannot be uniformly applied. Lightweight, deterministic mechanisms are required in time-critical paths, while stronger or more complex protections can be applied elsewhere. This latency-aware security model is a key architectural implication.

Finally, the inclusion of user safety and social risks broadens the notion of secure system design. Threats such as harassment, impersonation, and manipulative interactions undermine trust and long-term participation. Consequently, MM systems require an integrated approach that combines technical safeguards with mechanisms promoting trust, transparency, and user agency.

Notably, the stakeholder survey involved a relatively small sample (14 experts) drawn exclusively from a single research consortium. Although this ensured broad coverage of the technological components considered in this work, it may also introduce selection and sampling bias, potentially increasing consensus among participants while underrepresenting perspectives from independent security researchers, practitioners outside the consortium, or end users. Consequently, the prioritization of risks should be interpreted as an expert-informed assessment rather than a statistically representative characterization of the broader MM community.

\subsection{Design implications} 

Based on the results, we derive a set of design guidelines for secure and privacy-aware MM systems that balance strong protection with strict real-time constraints and the expressive nature of musical interaction.

\textbf{Latency-Aware Security Design:} 
Security mechanisms must be explicitly adapted to latency constraints. Ultra-low-latency interaction requires lightweight, deterministic, and stream-oriented protection (e.g., SRTP-like approaches), avoiding buffering, retransmissions, or complex handshakes. Less time-critical paths can accommodate stronger or more scalable mechanisms.

\textbf{Differentiation of Interaction Paths:}
A uniform security layer is insufficient. Systems should distinguish between latency-critical and non-critical communication paths, separating real-time interaction from control, signaling, and analytics. This enables stronger protection where latency constraints are less stringent.

\textbf{Data Minimization and Purpose Limitation:}
Given the high sensitivity of expressive and biometric data, systems should adopt data minimization and purpose limitation by design. Data should be processed only when necessary and protected according to its sensitivity level, as different data types pose different risks.

\textbf{Local and Edge-Centric Processing:}
Sensitive data (particularly biometric and expressive streams) should be processed at the edge or locally whenever possible. This reduces exposure, limits aggregation risks, and aligns with real-time performance requirements.

\textbf{Protection of Expressive and Creative Content:}
Security mechanisms must explicitly protect artistic content during creation and interaction. This includes preventing unauthorized recording, manipulation, and redistribution, as well as preserving integrity and attribution in real time.

\textbf{Robust Identity and Access Control:}
Identity management must support dynamic participation while preventing impersonation. This requires strong authentication, context-aware authorization, and protection of avatar-based identities across heterogeneous environments.

\textbf{Protection Against Inference and Re-Identification:}
Design must address inference risks arising from expressive and behavioural data. Even anonymized data may enable re-identification; therefore, aggregation, access, and AI processing pipelines must be carefully controlled.

\textbf{Performance-Aware Security Placement:}
Security controls should be strategically distributed. Lightweight mechanisms should be used in latency-critical paths, while heavier protections can be applied in storage, analytics, and non-real-time components.

\textbf{Transparency, User Awareness, and Control:}
Users must be informed about data collection and processing, especially for biometric sensing. Systems should provide clear consent mechanisms, visibility, and control over data sharing to maintain trust.

\textbf{Security-by-Design and Explicit Trust Boundaries:}
Security and privacy must be integrated from the outset. Clear trust boundaries, minimal implicit trust, and explicit architectural assumptions improve resilience in distributed MM systems.

\section{Conclusions} 

This paper has presented a structured analysis of security and privacy challenges in the MM, showing how its combination of ultra-low-latency interaction, continuous multimodal sensing, heterogeneous infrastructures, and real-time creative collaboration creates a distinctive risk landscape. Although MM systems inherit many threats from the broader Metaverse, their impact is amplified by the sensitivity of timing, expressive interaction, and IP.

The analysis identified the main assets requiring protection, including live musical content, expressive interaction data, identity and session metadata, IP, and service availability, and showed that the most critical risks are neurophysiological data leakage and real-time stream disruption. These findings demonstrate that privacy, availability, and interaction quality are tightly coupled in MM environments. Furthermore, the proposed layered threat model shows that vulnerabilities span network, application, data/AI, device, IPR, and social dimensions, requiring a holistic, cross-layer security approach.

A key contribution of this work is the evaluation of existing security protocols under strict latency constraints. The results indicate that conventional approaches (e.g., TLS over TCP) are generally unsuitable for real-time musical interaction, whereas lightweight, stream-oriented mechanisms such as SRTP and DTLS provide a better balance between protection and performance. More broadly, no single protocol or architectural solution is sufficient across all MM scenarios; instead, security must be adapted to latency constraints, data sensitivity, and interaction contexts.

Beyond protocol selection, this work highlights the importance of adopting security-by-design and privacy-by-design principles. Effective MM systems should integrate latency-aware protection strategies, data minimization, edge-centric processing, explicit trust boundaries, and user awareness mechanisms into their architecture.

Several open challenges remain, including the governance of biometric and neurophysiological data, interoperable enforcement of IP rights across platforms, and scalable trust management in distributed, multi-stakeholder ecosystems. Addressing these challenges will be essential for enabling secure, trustworthy, and sustainable MM environments. 

Ultimately, this work demonstrates that security and privacy are not auxiliary concerns, but fundamental requirements for meaningful, high-quality, and trustworthy musical interaction.


\bibliographystyle{IEEEtran}
\bibliography{article_specific.bib,IoMusT.bib,musical_XR_and_MM.bib,NMP.bib,Turchet.bib}


\end{document}